\documentclass[UTF8,aps,prc,showpacs,twocolumn,superscriptaddress]{revtex4}
\usepackage{multirow}
\usepackage{graphicx}
\usepackage{amsmath}
\usepackage{mathrsfs}
\usepackage{ulem}
\usepackage{threeparttable}
\usepackage{rotating}
\usepackage{diagbox}

\def\version{version}

\newcommand{ \be }{\begin{equation}}
\newcommand{ \ee }{\end{equation}}
\newcommand{ \bea }{\begin{eqnarray}}
\newcommand{ \eea }{\end{eqnarray}}

\usepackage{color}

\begin{document}
\title{
\begin{flushright}
{
\small \sl \version 1.3\\
}
\end{flushright}

Factorization of event-plane correlations over transverse momentum in relativistic heavy ion collisions in a multi-phase transport model

}

\author{Kai Xiao}\email{kxiao@mails.ccnu.edu.cn}
\affiliation{College of Electronics and Information Engineering, South-Central University for Nationalities, Wuhan 430074, China}

\author{Feng Liu}\email{fliu@mail.ccnu.edu.cn}
\affiliation{Key Laboratory of Quark and Lepton Physics (MOE) and Institute of Particle Physics, Central China Normal University, Wuhan 430079, China}

\author{Fuqiang Wang}\email{fqwang@purdue.edu}
\affiliation{Key Laboratory of Quark and Lepton Physics (MOE) and Institute of Particle Physics, Central China Normal University, Wuhan 430079, China}
\affiliation{Department of Physics, Purdue University, West Lafayette, Indiana 47907, USA}

\date{\today}

\begin{abstract}
Momentum-space azimuthal harmonic event planes (EP) are constructed from final-state mid-rapidity particles binned in transverse momentum ($p_T$) in $\sqrt{s_{NN}}$ = 200 GeV Au+Au collisions in a multi-phase transport (AMPT) model. The EP correlations between $p_T$ bins, corrected by EP resolutions, are smaller than unity. This indicates that the EP's decorrelate over $p_T$ in AMPT, qualitatively consistent with data and hydrodynamic calculations. It is further found that the EP correlations approximately factorize into single $p_T$-bin EP correlations to a common plane. This common plane appears to be the momentum-space EP integrated over all $p_T$, not the configuration-space participant plane (PP).

\end{abstract}

\maketitle
\clearpage
\section{INTRODUCTION}

In high energy heavy ion collisions, the initial high energy density and pressure buildup drive the collision system to rapid expansion. In non-central collisions, the pressure gradient is anisotropic due to the anisotropic transverse overlap geometry, resulting in an observable momentum-space azimuthal anisotropy~\cite{Ollitrault}. The particle azimuthal distribution can be expressed in Fourier series:

\begin{equation}
\frac{d^2N}{dp_{T}d\phi} \propto 1 + \sum 2v_{n}\cos [n(\phi - \Psi_{n})],
\end{equation} where $\phi$ represents the particle azimuthal angle and $\Psi_{n}$ is the $n^{th}$ harmonic plane angle. The Fourier coefficients~\cite{Voloshin}

\begin{equation}
v_{n} = \langle\cos[n(\phi - \Psi_{n})] \rangle,
\end{equation} characterize the magnitudes of the harmonic anisotropies, where the bracket $\langle...\rangle$ denotes averages over particles and events. In the limit of smooth nuclei, odd harmonics vanish; $\Psi_{2}$ coincides with the reaction plane (the plane defined by the beam direction and the impact parameter vector), and elliptic flow ($v_{2}$) is the leading component because of the predominant elliptic overlap shape. Large $v_{2}$ has been measured experimentally~\cite{whitepapers}. In fact, the measured $v_{2}$ is so large that hydrodynamic descriptions are applicable, and a very small shear viscosity to entropy density ratio ($\eta$/s) is required~\cite{Heinz}. This constitutes an important evidence for the formation of a new form of matter, the strongly interacting quark-gluon plasma (sQGP) in relativistic heavy ion collisions~\cite{whitepapers}. Due to fluctuations, the energy density in the overlap region is lumpy, so all harmonics can exist~\cite{Alver} and the harmonic planes of different orders are unnecessarily the same. Higher order harmonics are more sensitive to $\eta$/s, and hence play an important role in extracting precise information about the sQGP medium~\cite{Gombeaud}.

As the initial configuration space information of the overlap region is not experimentally accessible, the harmonic symmetry plane $\Psi^{PP}_{n}$ in configuration space, called the participant plane (PP), cannot be measured. In experiment, anisotropic flow is measured via final-state particle correlations~\cite{whitepapers}. For example, one constructs an event plane (EP), $\Psi^{EP}_{n}$, by the final-state particle momenta, as a proxy for $\Psi^{PP}_{n}$~\cite{Poskanzer}. One then correlates a test particle with the $\Psi^{EP}_{n}$ to measure anisotropic flow via Eq.(2) where the deviation of $\Psi^{EP}_{n}$ from $\Psi_{n}$ is corrected by a resolution factor~\cite{Poskanzer}. Alternatively, the anisotropic flow is measured using azimuthal correlations between the observed particles, where the Fourier coefficients of the correlation function are given by

\begin{equation}
V_{n}(p^{a}_{T},p^{b}_{T}) \equiv \langle \cos [n (\phi^{a} - \phi^{b})] \rangle.
\end{equation} Often factorization is assumed where the two-particle correlations arise only from the single particle correlations to a common harmonic plane $\Psi_{n}$ (i.e. nonflow intrinsic particle correlations are neglected). Under such conditions, since the common $\Psi_{n}$ cancel in ($\phi^{a}$ - $\Psi_{n}$) - ($\phi^{b}$ - $\Psi_{n}$), one has

\begin{equation}
V_{n}(p^{a}_{T},p^{b}_{T}) = v_{n}(p^{a}_{T})v_{n}(p^{b}_{T}).
\end{equation} The single-particle anisotropic flow $v_{n}$ can be simply obtained as the square root of the two-particle cumulant measurement in Eq.(3) by choosing the two particles from the same phase space, or when the $v_{n}$ of one particle is known (referred to as the reference particle).

Studies have shown, however, that the harmonic planes, whether in configuration or momentum space, are not the same at different pseudorapidities. Previous works~\cite{BJ,Xiao} have shown that EP's are decorrelated over pseudorapidity ($\eta$) and the decorrelation increases with increasing $\eta$ gap between the two particles.

Not only does the harmonic plane depend on $\eta$, but also on transverse momentum ($p_T$). Gardim et al.~\cite{Gardim} have pointed out with ideal hydrodynamic calculations that the fluctuating flow angles $\Psi_{n}$ depend on $p_T$. Viscous hydrodynamic calculations~\cite{HQS} indicate that the harmonic planes decorrelate over $p_T$ even in the same $\eta$ region.  This is indeed observed by CMS in Pb+Pb and p+Pb collisions at the LHC~\cite{CMS}. The experimental confirmation of the $p_{T}$ decorrelation is taken as a strong evidence supporting hydrodynamic descriptions of relativistic heavy ion collisions.

Experimentally, in order to reduce nonflow correlations (such as small angle jet correlations and resonance decays), one often apply $\eta$ gap between the two particles used in correlation measurements. In measuring high $p_{T}$ particle anisotropy, the reference particle is often taken at low $p_{T}$. The harmonic plane decorrelation over $\eta$ and $p_{T}$ casts a problem in those measurements; it may require the EP to be constructed in the same phase space as the particle of interest in order to measure flow anisotropy~\cite{Xiao}. As an operational definition, one may define the flow of particles in a particular phase space as that with respect to the harmonic plane of the same phase space region. Namely,

\begin{equation}
v_{n}(\eta, p_{T}) = \langle \cos [n(\phi(\eta, p_{T}) - \Psi_{n}(\eta, p_{T}))]\rangle.
\end{equation} We note, however, that such $v_n$ is, in realty, contaminated by non-negligible nonflow effects.

The $\eta$ and $p_{T}$ decorrelation of the harmonic planes explicitly breaks factorization of Eq.(4). This is easy to see. The two-particle cumulant measurement with the two particles coming from different phase spaces is

\begin{equation}
\begin{split}
V_{n}(p^{a}_{T},p^{b}_{T}) & \equiv \langle\cos [n (\phi_{n}(p^{a}_{T}) - \phi_{n}(p^{b}_{T}))]\rangle \\
& = v_{n}(p^{a}_{T})v_{n}(p^{b}_{T})\cos [n (\Psi_{n}(p^{a}_{T}) - \Psi_{n}(p^{b}_{T}))].
 \end{split}
\end{equation} The two harmonic planes from different phase spaces are now different and can no longer cancel in two-particle azimuthal correlations. Note that in Eq.(6) we have neglected possible nonflow contributions (which is usually small when a large phase-space gap is imposed between the two particles), so that we can write the average of the product into the product of averages.

The EP decorrelations were experimentally measured to be appreciable~\cite{CMS}. However, the experimentally measured two-particle cumulants obey factorization exceedingly well~\cite{Factor}. It implies that $\cos [n (\Psi_{n}(p^{a}_{T}) - \Psi_{n}(p^{b}_{T}))]$ correlations between two phase spaces, $a$ and $b$, must factorize to a good degree~\cite{fqw}. This is by no means obvious.

In this paper we try to address two questions. One, are EP's in parton transport model similarly decorrelated as in data and hydrodynamic calculations? Second, how well are the EP correlations factorized? The answers to these questions can hopefully reveal additional insights to the physics of EP decorrelations.

\begin{table*}[]
\renewcommand\arraystretch{2.}
\centering
\scriptsize
\begin{tabular}{|c|c|c|c|c|c|}
\hline
$p_{T}$(GeV/c) & $n=2$ & $n=3$  & $p_{T}$(GeV/c) & $n=2$ & $n=3$\\\hline

$0-0.2$ & $0.11\pm0.03$ & $0.06\pm0.04$  & $0.2-2.0$ & $0.889\pm0.001$ & $0.568\pm0.004$\\\hline

$0.2-0.4$ & $0.557\pm0.005$ & $0.17\pm0.02$  & $0-0.2~\|~0.4-2.0$ & $0.853\pm0.002$ & $0.537\pm0.004$\\\hline

$0.4-0.6$ & $0.719\pm0.003$ & $0.339\pm0.008$  & $0-0.4~\|~0.6-2.0$ & $0.810\pm0.002$ & $0.435\pm0.006$\\\hline

$0.6-0.8$ & $0.684\pm0.003$ & $0.384\pm0.007$  & $0-0.6~\|~0.8-2.0$ & $0.829\pm0.002$ & $0.435\pm0.006$\\\hline

$0.8-1.0$ & $0.608\pm0.004$ & $0.364\pm0.007$  & $0-0.8~\|~1.0-2.0$ & $0.847\pm0.002$ & $0.462\pm0.005$\\\hline

$1.0-1.2$ & $0.525\pm0.006$ & $0.313\pm0.009$  & $0-1.0~\|~1.2-2.0$ & $0.858\pm0.001$ & $0.480\pm0.005$\\\hline

$1.2-1.4$ & $0.423\pm0.008$ & $0.26\pm0.01$  & $0-1.2~\|~1.4-2.0$ & $0.864\pm0.001$ & $0.496\pm0.005$\\\hline

$1.4-1.6$ & $0.36\pm0.01$ & $0.23\pm0.01$  & $0-1.4~\|~1.6-2.0$ & $0.866\pm0.001$ & $0.506\pm0.005$\\\hline

$1.6-1.8$ & $0.32\pm0.01$ & $0.21\pm0.01$  & $0-1.6~\|~1.8-2.0$ & $0.867\pm0.001$ & $0.507\pm0.005$\\\hline

$1.8-2.0$ & $0.33\pm0.01$ & $0.23\pm0.01$  & $0-1.8$ & $0.867\pm0.001$ & $0.507\pm0.005$\\\hline

\hline
\end{tabular}
 \vspace{1em}
\caption{(Color online) The resolutions, $\Re^{EP}_{n}$ of the second- and third-order harmonic EP's constructed by final-state particle momentum azimuthal angles in different transverse momentum ($p_{T}$) bins. Particles are taken from pseudo-rapidity range of $|\eta| < 1.0$ in 20-40\% centrality (b = 6.79-9.61 fm) Au+Au collisions at $\sqrt{s_{NN}} = 200$ GeV simulated by the AMPT model (string melting, 3 mb parton cross section).}
\vspace{3em}
\end{table*}

\begin{table*}[]
\renewcommand\arraystretch{2.}
\centering
\scriptsize
\begin{tabular}{|c|c|c|c|c|c|c|c|c|c|c|}

    \hline
     \diagbox{$p^{b}_{T}$}{$p^{a}_{T}$} & $0-0.2$ & $0.2-0.4$ & $0.4-0.6$ & $0.6-0.8$ & $0.8-1.0$ & $1.0-1.2$ & $1.2-1.4$ & $1.4-1.6$ & $1.6-1.8$ & $1.8-2.0$ \\\hline

     $1.8-2.0$ & $-0.32\pm0.38$ & $0.79\pm0.14$ & $0.70\pm0.07$ & $0.82\pm0.06$ & $0.86\pm0.07$ & $0.92\pm0.08$ & $0.90\pm0.09$ & $0.70\pm0.12$ & $0.65\pm0.12$ & ~~~~~~~~~~~~~~~~\\\hline

     $1.6-1.8$ &$-0.43\pm0.43$ & $0.85\pm0.16$ & $0.87\pm0.08$ & $0.91\pm0.07$ & $0.77\pm0.08$ & $0.92\pm0.09$ & $0.81\pm0.11$ & $0.98\pm0.13$ & ~~~~~~~~~~~~~~~~ & $0.86\pm0.05$\\\hline

     $1.4-1.6$ & $-0.79\pm0.43$ & $0.51\pm0.16$ & $1.00\pm0.08$ & $1.07\pm0.07$ & $0.98\pm0.08$ & $1.10\pm0.09$ & $1.06\pm0.11$ & ~~~~~~~~~~~~~~~~ & $0.94\pm0.05$ & $0.90 \pm0.05$\\\hline

     $1.2-1.4$ & $0.16\pm0.34$ & $0.66\pm0.13$ & $1.00\pm0.06$ & $1.00\pm0.06$ & $1.06\pm0.06$ & $0.87\pm0.07$ & ~~~~~~~~~~~~~~~~ & $0.97\pm0.04$ & $0.93\pm0.04$ & $0.92\pm0.04$\\\hline

     $1.0-1.2$ &$-0.32\pm0.28$ & $0.88\pm0.11$ & $0.85\pm0.05$ & $0.88\pm0.05$ & $1.03\pm0.05$ & ~~~~~~~~~~~~~~~~ & $1.02\pm0.02$ & $0.95\pm0.03$ & $0.88\pm0.03$ & $0.85\pm0.03$\\\hline

     $0.8-1.0$ &$-0.32\pm0.24$ & $0.68\pm0.09$ & $1.00\pm0.05$ & $0.88\pm0.04$ & ~~~~~~~~~~~~~~~~ & $0.98\pm0.02$ & $0.95\pm0.02$ & $0.97\pm0.03$ & $0.93\pm0.03$ & $0.86\pm0.03$\\\hline

     $0.6-0.8$ &$-1.00\pm0.23$ & $0.82\pm0.09$ & $0.88\pm0.04$ & ~~~~~~~~~~~~~~~~ & $0.95\pm0.01$ & $0.97\pm0.01$ & $0.96\pm0.02$ & $0.92\pm0.02$ & $0.93\pm0.03$ & $0.89\pm0.02$\\\hline

     $0.4-0.6$ &$-0.45\pm0.26$ & $0.76\pm0.10$ & ~~~~~~~~~~~~~~~~ & $0.97\pm0.01$ & $0.96\pm0.01$ & $0.95\pm0.01$ & $0.94\pm0.02$ & $0.90\pm0.02$ & $0.91\pm0.02$ & $0.86\pm0.02$\\\hline

     $0.2-0.4$ &$0.60\pm0.51$ & ~~~~~~~~~~~~~~~~ & $0.95\pm0.01$ & $0.94\pm0.01$ & $0.95\pm0.02$ & $0.97\pm0.02$ & $0.93\pm0.02$ & $0.92\pm0.03$ & $0.96\pm0.03$ & $0.87\pm0.03$\\\hline

     $0-0.2$ & ~~~~~~~~~~~~~~~~ & $0.81\pm0.09$ & $0.97\pm0.07$ & $0.87\pm0.08$ & $0.92\pm0.09$ & $1.06\pm0.10$ & $1.00\pm0.12$ & $0.94\pm0.14$ & $0.99\pm0.16$ & $0.86\pm0.16$\\\hline

  \end{tabular}

  \vspace{1em}
  \caption{(Color online) The resolution-corrected EP correlation strengths $C_{2}(p^{a}_{T},p^{b}_{T})$ (right down corner) and $C_{3}(p^{a}_{T},p^{b}_{T})$ (left up corner) of Eq.(9) for different combinations of $p^{a}_{T}$ and $p^{b}_{T}$ ($p^{a}_{T} \neq p^{b}_{T}$). Particles used in EP construction are restricted within pseudo-rapidity $|\eta| < 1.0$. The data are simulated by AMPT (string melting, 3 mb cross section) in 20-40\% centrality (b = 6.79-9.61 fm) in Au+Au collisions at $\sqrt{s_{NN}} = 200$ GeV.}
  \end{table*}

\section{Analysis Method}

\begin{figure*}[htb]
\vskip 0.cm
\centerline{\includegraphics[width=0.85\textwidth]{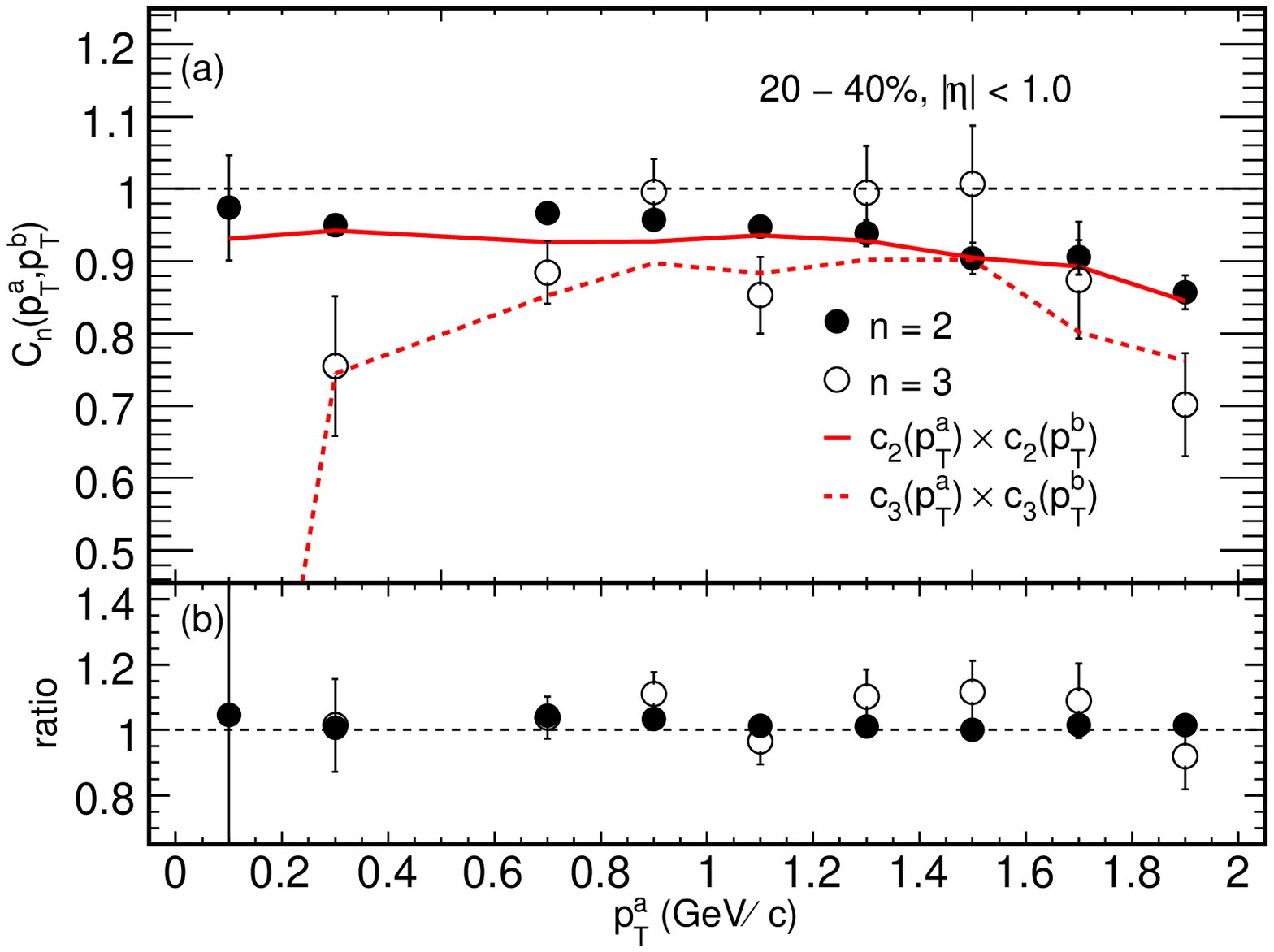}}
\vspace{-8em}
\caption{(Color online) Upper panel shows the EP correlation strength $C_{n}(p^{a}_{T},p^{b}_{T})$, corrected by the corresponding EP resolutions, as a function of $p^{a}_{T}$ for a fixed $p^{b}_{T}$ bin. Particles used in EP construction are restricted within pseudo-rapidity $|\eta| < 1.0$. The data are simulated by AMPT (string melting, 3 mb cross section) in 20-40\% centrality (b = 6.79-9.61 fm) in Au+Au collisions at $\sqrt{s_{NN}} = 200$ GeV. Red lines are the fit to data by the function of $c_{n}(p^{a}_{T})\times c_{n}(p^{b}_{T})$ where $c_{n}(p_{T})$ is a set of ten fit parameters. Lower panel shows the ratio of the data points to the fit result.}
\vspace{-1em}
\end{figure*}

The AMPT (A Multi-Phase Transport) model with string melting describes many experimental data reasonably well, particularly the anisotropic flow measurements~\cite{AMPTflow}. We thus use the AMPT model with string melting for our study. There are four main components in AMPT: the initial conditions, parton interactions, hadronization, and hadron interactions. The initial conditions are obtained from the HIJING model~\cite{AMPTini}, which includes the spatial and momentum information of minijet partons from hard processes and strings from soft processes. The time evolution of partons is then treated according to the ZPC parton cascade model~\cite{AMPTzpc}. After parton interactions cease, a combined coalescence and string fragmentation model is used for the hadronization of partons. The scattering among the resulting hadrons is described by a relativistic transport (ART) model~\cite{AMPTart} which includes baryon-baryon, baryon-meson and meson-meson elastic, and inelastic scatterings.

\begin{figure*}[htb]
\vskip 0.cm
\centerline{\includegraphics[width=0.85\textwidth]{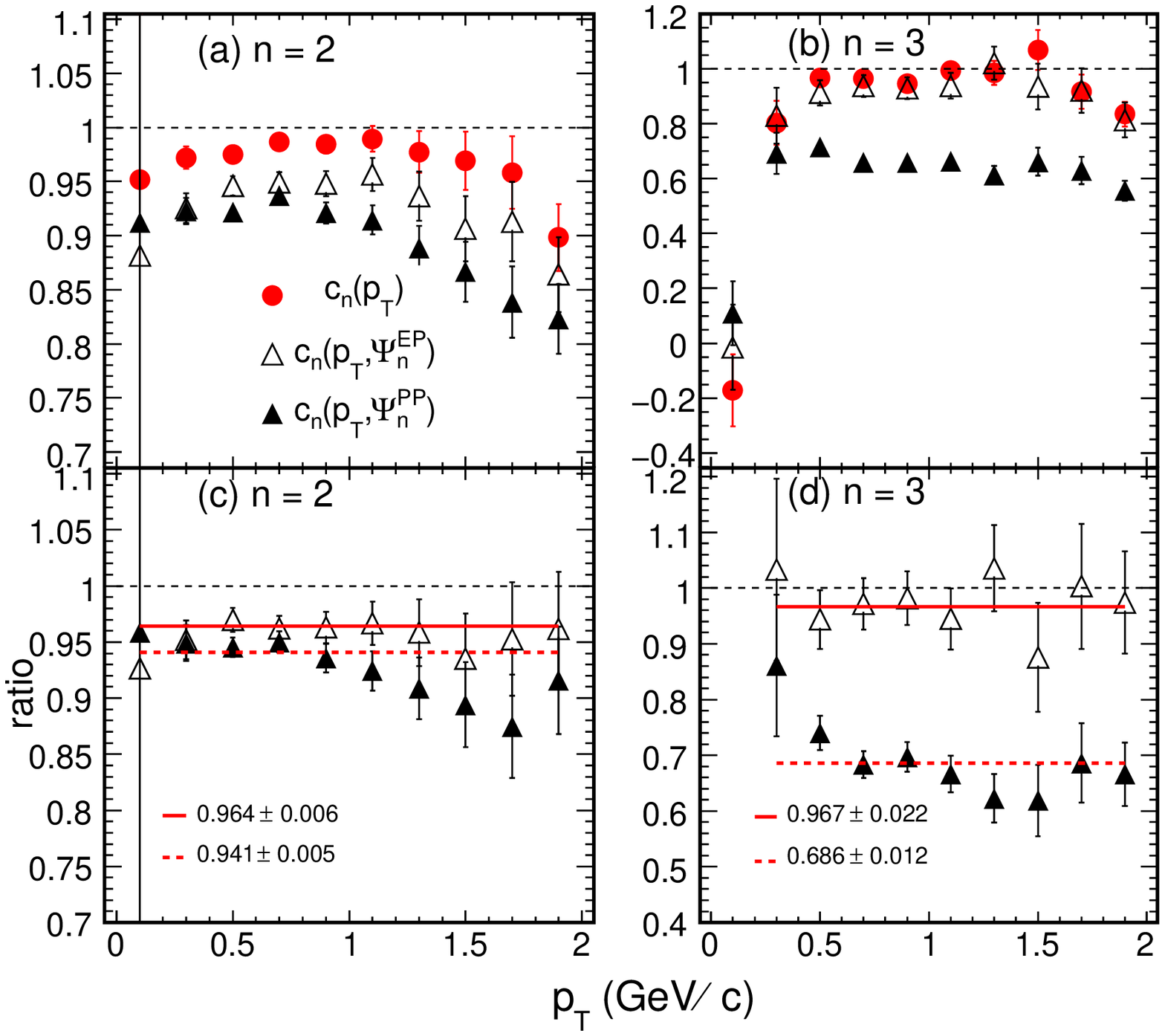}}
\vspace{-3em}
\caption{(Color online) Upper panels show the comparisons of the fit parameter $c_{n}(p_{T})$ to $c_{n}(p_{T},\Psi^{EP}_{n})$, the correlations between $\Psi^{EP}_{n}(p_{T})$ and the EP constructed from all $p_{T}$ (but excluding the relevant $p_{T}$ bin) and $c_{n}(p_{T},\Psi^{PP}_{n})$, the correlation between $\Psi^{EP}_{n}(p_{T})$ and the PP, for (a) second- and (b) third-order harmonic. Lower panels show the corresponding ratios.  Data are for 20-40\% centrality (b = 6.79-9.61 fm) in Au+Au collisions at $\sqrt{s_{NN}} = 200$ GeV from the AMPT model (string melting, 3 mb parton cross section).}
\vspace{-1em}
\end{figure*}

In AMPT model, two harmonic planes can be constructed. One is calculated from the initial configuration space information of partons by~\cite{Alver}

\begin{equation}
\Psi_{n}^{PP} = \frac{\rm atan2(\langle r^{n}\sin(n\phi_{r}) \rangle, \langle r^{n}\cos(n\phi_{r}) \rangle) + \pi}{n},
\end{equation} where $r$ and $\phi_{r}$ are the polar coordinate position of each parton. Because of event-by-event geometry fluctuations,  $\Psi^{PP}_{n}$ is not necessarily as same as the reaction plane. Experimentally, the coordinate space information is not accessible. Instead, the EP can be constructed from final-state particle momenta by

\begin{equation}
\Psi_{n}^{EP} = \frac{\rm atan2(\langle \sin(n\phi) \rangle, \langle \cos(n\phi) \rangle)}{n},
\end{equation} where $\phi$ is the azimuthal angle of the particle momentum.
Due to the finite multiplicity of constituents, the constructed harmonic plane is smeared from the true one, $\Psi^{true}_{n}$ --the geometry harmonic plane of the participant partons in configuration space in the limit of infinite parton multiplicity-- by a resolution factor. The resolution factor $\Re_{n} = \langle \cos [n (\Psi_{n} - \Psi^{true}_{n})]\rangle$ is calculated with an iterative procedure by the sub-event method, dividing the constituents randomly into two sub-events~\cite{Poskanzer}. Because of the large initial parton multiplicity, the calculated initial configuration-space PP resolution, $\Re^{PP}_{n}$, is nearly unity~\cite{Xiao}. However, the final-state hadron multiplicity is smaller so the final-state momentum-space EP resolution, $\Re^{EP}_{n}$, is smaller than unity. The EP resolutions in different $p_{T}$ bins are tabulated in Table I for both the second- and third-order harmonic EP's. In addition, the resolutions of the EP's constructed from all particles in $0<p_T<2$ GeV/c but excluding those in a particular $p_{T}$ bin are also listed.

\section{Results and Discussions}

The EP angles $\Psi^{EP}_{n}(p_{T})$ are constructed from final-state charged particle momenta in ten equal-width $p_{T}$ bins in $0<p_T<2$ GeV/c. Particles used in the EP construction are restricted within $|\eta|<1.0$. Both the elliptic ($n=2$) and triangular ($n=3$) EP's are constructed. The EP correlation between $p^{a}_{T}$ and $p^{b}_{T}$ bins, corrected by the EP resolutions in Table I, is given by:

\begin{equation}
C_{n}(p^{a}_{T}, p^{b}_{T}) = \frac{\langle\cos [n (\Psi^{EP}_{n}(p^{a}_{T}) - \Psi^{EP}_{n}(p^{b}_{T}))]\rangle}{\Re^{EP}_{n}(p^{a}_{T})\Re^{EP}_{n}(p^{b}_{T})}.
\end{equation} The EP correlation strengths $C_{n}$ are tabulated in Table II for both $n=2$ and $n=3$. For each $C_{n}(p^a_{T},p^b_{T})$ value in the table,we have already taken the average of $C_{n}(p^a_{T},p^b_{T})$ and $C_{n}(p^b_{T},p^a_{T})$. As seen from the table,the measured $C_{n}(p^{a}_{T}, p^{b}_{T})$ for $p^{a}_{T}\neq p^{b}_{T}$ are less than unity. This indicates that the EP's in AMPT are indeed decorrelated over $p_{T}$. The decorrelations are similar in magnitude to those observed in data at the LHC~\cite{CMS} and from hydrodynamic calculations~\cite{HQS}. Figure 1(a) shows the EP correlation strengths $C_{n}$ as a function of $p^{a}_{T}$ for a fixed $p^{b}_{T} = 0.4<p_{T}<0.6$ GeV/c, as an example.

To test factorizability, the EP correlations of all combinations of $p^{a}_{T}$ and $p^{b}_{T}$ are fitted by a set of ten parameter $c_{n}(p_{T})$ corresponding to the ten $p_{T}$ bins:

\begin{equation}
C_{n}(p^{a}_{T}, p^{b}_{T}) = c_{n}(p^{a}_{T})\times c_{n}(p^{b}_{T}).
\end{equation} Factorization of the EP correlations can be tested by the fitting quality. The fit result, $c_{n}(p^{a}_{T})\times c_{n}(p^{b}_{T})$, is shown as the curves in Fig.1 along with the EP correlations data points from AMPT, $C_{n}(p^{a}_{T}, p^{b}_{T})$. Fits are good for both n=2 and n=3. To illustrate the fit quality, Fig. 1(b) shows the ratio of the AMPT data to the fit. The ratio is consistent with unity. Note Fig.1 is only a subset of the EP correlation data for a fixed $p^{b}_{T}$ bin as an example, but all $10 \times 9/2$ correlation data for a given harmonic ($n=2$ or 3) are fitted simultaneously with ten fit parameters in a single fit. The fit $\chi^2$/ndf are 30.5/35 and 45.5/35 for $C_{2}$ and $C_{3}$, respectively. Our fit results indicate that the EP correlations are well factorized. The EP correlation between two $p_{T}$ bins is determined by only the properties of the two single-$p_{T}$ bins. This may indicate that the EP decorrelation is of a random nature.

Since $C_{n}(p^{a}_{T}, p^{b}_{T})$ factorize well into products of single $p_{T}$-bin EP quantities, the fit results suggest that the EP correlations may be caused by single $p_{T}$-bin EP correlations to a common plane, $\Psi_n$. Namely,

\begin{widetext}
\begin{equation}
\begin{split}
\frac{\langle\cos [n (\Psi^{EP}_{n}(p^{a}_{T}) - \Psi^{EP}_{n}(p^{b}_{T}))]\rangle}{\Re^{EP}_{n}(p^{a}_{T})\Re^{EP}_{n}(p^{a}_{T})}  & = \frac{\langle\cos [n (\Psi^{EP}_{n}(p^{a}_{T}) - \Psi_{n})]\rangle}{\Re^{EP}_{n}(p^{a}_{T})}
\times \frac{\langle\cos [n (\Psi^{EP}_{n}(p^{b}_{T}) - \Psi_{n})]\rangle}{\Re^{EP}_{n}(p^{b}_{T})},
\end{split}
\end{equation}
\end{widetext}

and

\begin{equation}
c_{n}(p_{T}) = \frac{\langle\cos [n (\Psi^{EP}_{n}(p_{T}) - \Psi_{n})]\rangle}{\Re^{EP}_{n}(p_{T})}.
\end{equation}

The good fitting quality demands an answer to the following question: what plane is $\Psi_{n}$? To address this question, we show in upper panel of Fig. 2 the comparisons between the fit parameter $c_{n}(p_{T})$ and that given by Eq.(12) where $\Psi_{n}$ is in one case calculated by the configuration-space $\Psi^{PP}_{n}$ and in the other case by the momentum-space $\Psi^{EP}_{n}$ from particles over the entire $p_{T}$ range:

\begin{equation}
c_{n}(p_{T},\Psi^{PP}_{n}) = \frac{\langle\cos [n (\Psi^{EP}_{n}(p_{T}) - \Psi^{PP}_{n})]\rangle}{\Re^{EP}_{n}(p_{T})\Re^{PP}_{n}}
\end{equation}

and

\begin{widetext}
\begin{equation}
\begin{split}
c_{n}(p_{T},\Psi^{EP}_{n}) =
\frac{\langle\cos [n (\Psi^{EP}_{n}(p_{T})-\Psi^{EP}_{n}(0<p_{T}<2 ~\text{GeV/c} \; \text{but exclude} \; p_{T}))]\rangle}{\Re^{EP}_{n}(p_{T})\Re^{EP}_{n}(0<p_{T}<2 ~\text{GeV/c} \; \text{but exclude} \; p_{T})}
\end{split}
\end{equation}
\end{widetext}respectively. The $\Psi^{PP}_{n}$ are constructed from the initial parton configuration within the pseudo-rapidity of $|\eta_{parton}| < 1.0$.  The $\Psi^{EP}_{n}$ are constructed from the final-state particle momenta within $|\eta_{hadron}| < 1.0$ and $0<p_{T}<2$ GeV/c but excluding the $p_{T}$ bin of interest. To remove auto-correlations, the construction of $\Psi^{EP}_{n}$ excludes particles of the particular $p_{T}$ bin used for the construction of $\Psi_{n}(p_{T})$. The resolutions $\Re^{PP}$ and $\Re^{EP}_{n}$($0<p_{T}<2$ GeV/c  but exclude $p_{T}$) are to take into account the inaccuracy of the constructed PP and EP (although $\Re^{PP}$ is found to be approximately unity).

The comparisons of $c_{n}(p_{T})$ to $c_{n}(p_{T},\Psi^{PP}_{n})$ and $c_{n}(p_{T},\Psi^{EP}_{n})$ are shown in Fig.2 for both the second- and third-order harmonic EP's. We find that $c_{n}(p_{T},\Psi^{EP}_{n})$ is more consistent with the fit parameter $c_{n}(p_{T})$ than $c_{n}(p_{T},\Psi^{PP}_{n})$ is. The more quantitative comparisons are shown in Fig. 2(c) and (d) where the ratios are depicted. The ratio of $c_{3}(p_{T},\Psi^{EP}_{3})$/$c_{3}(p_{T})$ is consistent with unity while $c_{3}(p_{T},\Psi^{PP}_{3})$/$c_{3}(p_{T})$ is not. For $n=2$, both ratios are smaller than unity, however, $c_{2}(p_{T},\Psi^{EP}_{2})$/$c_{2}(p_{T})$ is closer to unity than $c_{2}(p_{T},\Psi^{PP}_{2})$/$c_{2}(p_{T})$. The deviation of $c_{2}(p_{T},\Psi^{EP}_{2})$/$c_{2}(p_{T})$ from unity could be due to nonflow effect where some particles are intrinsically correlated in the final-state momentum space. Our results show that the global harmonic plane is not the $\Psi^{PP}_{n}$; it may be the $\Psi^{EP}_{n}$. In other words, $\Psi^{EP}_{n}(p_{T})$ fluctuates randomly around $\Psi^{EP}_{n}$, not $\Psi^{PP}_{n}$.

\vspace{3cm}
\section{Conclusions}

In summary, we have studied momentum-space event plane (EP) correlations in b = 6.79-9.61 fm Au+Au collisions at $\sqrt{s_{NN}}$ = 200 GeV using the AMPT model (string melting, 3 mb parton cross section). The EP's are constructed from final-state mid-rapidity particles binned in $p_{T}$. Both the second- and third-order harmonic EP's are studied. The EP correlations between $p_{T}$ bins are corrected by the corresponding EP resolutions. The EP correlations are found to be smaller than unity, indicating that the EP's decorrelate over $p_{T}$ in AMPT, qualitatively consistent with data and hydrodynamic calculations. It is further found that the EP correlations factorize into single $p_{T}$-bin EP correlations to a common plane. This common plane appears to be approximately the momentum-space EP integrated over all $p_{T}$, but not the configuration-space participant plane (PP).


\section{Acknowledgments}
This work is supported in part by MOST of China under 973 Grant 2015CB856901, the National Natural Science Foundation of China under grant No. 11228513, 11221504 and 11135011, U.S. Department of Energy under Grant No. DE-FG02-88ER40412, Fundamental Research Funds for the Central Universities (Grant No. CZQ15001).

%

%
\end{document}